\documentclass[twocolumn,english,prb,aps,showpacs]{revtex4}

\usepackage{color} 

\usepackage{graphicx}
\usepackage{bm}
\usepackage{amsmath}
\usepackage[dvipdfm,bookmarks=true,bookmarksnumbered=true,bookmarkstype=toc]{hyperref}

\begin{document}

\title{Microscopic calculation of thermally-induced spin-transfer torques 
} 
\author{Hiroshi Kohno$^1$, Yuuki Hiraoka$^2$, Moosa Hatami$^3$ and Gerrit E.~W. Bauer$^{3,4}$}
\affiliation{$^1$Department of Physics, Nagoya University, Furo-cho, Chikusa-ku, Nagoya, 464-8601, Japan 
\\
$^2$Graduate School of Engineering Science, Osaka University, Toyonaka, Osaka 560-8531, Japan 
\\
$^3$Kavli Institute of NanoScience, Delft University of Technology, Lorentzweg 1, 2628 CJ Delft, The Netherlands 
\\
$^4$Institute for Materials Research and WPI-AIMR, Tohoku University, Sendai 980-8577, Japan}
\date{\today}

\begin{abstract}
 Spin-transfer torques induced by temperature gradients in conducting ferromagnets 
are calculated microscopically for smooth magnetization textures. 
 Temperature gradients are treated {\it \`a la} Luttinger by introducing a fictitious gravitational field that couples to the energy density. 
 The thermal torque coefficients obtained by the Kubo formula contain  divergences caused by equilibrium components that should be subtracted before applying the Einstein-Luttinger relation. 
 Only by following this procedure a familiar Mott-like formula is obtained for the dissipative spin-transfer torque.  
 The result indicates that a fictitious field coupled to the entropy rather than energy would solve the issue from the outset. 

\end{abstract}
\maketitle

\section{Introduction}
\label{Introduction}

  A spin current is a flow of angular momentum, which can be transferred to other degrees of freedom and thereby exerts a torque on them. 
 In ferromagnetic conductors, an ordinary (Ohmic) electric current, induced by an applied electric field, 
is accompanied by a spin current, and this can be utilized to control magnetization dynamics.\cite{Slonczewski96, Berger96}  

 Spin currents can also be induced by a temperature gradient in ferromagnets, 
which may also be used to control magnetization without the need to apply an electric field.\cite{Berger85, Hatami07, Kovalev09} 
 A temperature gradient of $0.2$ K/nm in permalloy has been estimated to induce a torque equivalent to that by an electric 
current density of 10$^7$ A/cm$^2$.\cite{Bauer09}  
 This value indicates that thermally-driven domain wall motion may be realized in magnetic nanostructures. 
 Domain wall in magnetic insulators, in which spin currents are carried by magnons, 
indeed move under a temperature gradient.\cite{Jiang13} 

 In this paper, we calculate spin torques induced by a temperature gradient in a conducting ferromagnet focussing on mobile conduction electrons (not magnons). 
 We consider a general but smooth magnetization texture as described by the Landau-Lifshitz-Gilbert (LLG) equation. 
 To treat thermal perturbation, we follow Luttinger \cite{Luttinger64} and introduce a (fictitious) gravitational field 
which couples to the energy (or heat) density of the system and exploit the Einstein relation.\cite{Einstein05}   
 The calculation then should be done quite in parallel with the calculation of (ordinary) electrically-induced torques. 
 However, a straightforward calculation leads to a physically wrong contribution which diverges towards absolute zero temperature. 
 The resolution of this difficulty is the main subject of this paper.

 A similar problem has been known to exist in thermoelectric transport in strong magnetic fields.\cite{Oji85,Cooper97,Qin11} 
 In this case, the problem was resolved by separating the transport current from the magnetization current and applying the Einstein relation to the former. 
 In calculating spin torques, we need to generalize this idea and propose to separate the non-equilibrium and equilibrium components, applying the Einstein relation to the former. 
 A similar feature exists in the \lq spin-orbit torques' due to Rashba-type spin-orbit coupling, 
which will be reported in a separate paper.\cite{vdBijl} 

 This paper is organized as follows. 
 After a brief description of spin torques in Sec.~II, we define a model in Sec.~III. 
 Based on the formulation outlined in Sec.~IV, we evaluate explicitly the thermal torque in Sec.~V and observe that the result contains an unphysical contribution. 
 The resolution of this problem is described in Sec.~VI, and the correct result is given in Sec.~VII.  
 The consequence of our results are illustrated in Sec.~VIII for thermal torques in the absence of applied electric fields.
 In Sec.~IX, we discuss our procedure in a more general context. 
 The work is summarized in Sec.~X. 
 Some supplementary calculations and discussions are deferred to the Appendices.

\section{General Description of Spin Torques}
\label{General}

 The LLG-Slonczewski (LLGS) equation, in which the effects of spin currents are included, reads 
\begin{eqnarray}
  \dot {\bm n} = \gamma_0 {\bm H}_{\rm eff} \times {\bm n} 
  + \alpha_0 \dot {\bm n} \times {\bm n} + \tilde {\bm t} ,
\label{eq:LLG}
\end{eqnarray}
where ${\bm n} = {\bm n}({\bm r},t)$ is a unit-vector field representing 
the spin direction of magnetization, and the dot represents the time derivative. 
 The first two terms, a precessional torque 
(${\bm H}_{\rm eff}$: effective field, $\gamma_0$: gyromagnetic constant) 
and Gilbert damping ($\alpha_0$: damping constant), 
exist even without conduction electrons. 
 Effects of conduction electrons are given by the third term, 
$\tilde {\bm t}$, called spin torque in general. 

 For a smooth magnetization texture, ${\bm n}$, 
the torques due to an electrically-induced spin current density ${\bm j}_{\rm s} = {\bm j}_\uparrow - {\bm j}_\downarrow$ 
has the form, 
\begin{eqnarray}
 \tilde {\bm t}_{\rm el}
= - ({\bm v}_{\rm s} \!\cdot\! {\bm \nabla})\,  {\bm n} 
  - \beta \, {\bm n} \times ({\bm v}_{\rm s} \cdot\! {\bm \nabla}) 
    \, {\bm n} . 
\label{eq:t_el_0}
\end{eqnarray}
 The first term is the celebrated spin-transfer torque \cite{BJZ98}  
with the (renormalized) \lq\lq spin-transfer velocity'' 
\begin{eqnarray}
 {\bm v}_{\rm s} = - \frac{\hbar}{2e s_{\rm tot}} \, {\bm j}_{\rm s} , 
\label{eq:vs0}
\end{eqnarray}
where $s_{\rm tot}$ is the angular-momentum density 
of total magnetization (including conduction electrons, see Ref.~\onlinecite{Kohno08}). 
 The electron charge is denoted as $-e$ so that $e>0$. 
  The second so-called \lq $\beta$-term' originates from spin-relaxation processes.\cite{Zhang04,Thiaville05}  
 Although the dimensionless constant $\beta$ is expected to be small 
($\sim 0.01$), it importantly affects the dynamics of a domain wall.\cite{Zhang04,Thiaville05,TK04,Barnes05}

 Torques induced by a temperature gradient, $\nabla T$, take the same form 
\begin{eqnarray}
 \tilde {\bm t}_{\rm th} 
= - ({\bm v}_T \!\cdot\! {\bm \nabla})\,  {\bm n} 
  - \beta_T \, {\bm n} \times ({\bm v}_T \cdot\! {\bm \nabla}) 
    \, {\bm n} ,
\label{eq:t_th_0}
\end{eqnarray}
but the coefficient vector ${\bm v}_T$ is driven by $\nabla T$. 
 Any spin-relaxation process is expected to produce the second term, 
with $\beta_T$ being a dimensionless parameter. 
 By scattering theory Hals {\it et al.}\cite{Hals10} demonstrated that $\beta_T \ne \beta$, 
but a formulation by linear response theory is still lacking.

\section{Model}
\label{Model}

 The microscopic origin of spin torques is the $s$-$d$ exchange interaction\cite{com_sd}
\begin{eqnarray}
 H_{sd}  
= - M \int d^3 x \, {\bm n}(x) \!\cdot\! \hat {\bm \sigma} (x) ,
\label{eq:H_sd}
\end{eqnarray}
between the spin 
$\hat{\bm \sigma} (x) \equiv c^\dagger {\bm \sigma}\, c$ 
of conduction electrons and magnetization ${\bm n}(x)$, 
where $M$ is a coupling constant.  
 For example, if an electron moves through a magnetization texture 
${\bm n}(x)$, 
its spin experiences a time-dependent \lq field' $M{\bm n}$. 
 The electron, in turn, exerts a reaction (spin) torque \cite{Berger,TKS08} 
\begin{eqnarray}
 {\bm t}_{sd} = M {\bm n}(x) \times \langle \hat {\bm \sigma} (x) \rangle ,
\label{eq:t_sd_0}
\end{eqnarray}
on the magnetization since $M \langle \hat {\bm \sigma} (x) \rangle$ 
is an effective magnetic field,  where the brackets $\langle \cdots \rangle$ indicate a quantum statistical average. 
 The calculation of the torque is thus reduced to calculating the 
electron spin density in the current-carrying non-equilibrium state.

 To be specific, let us consider a free electron system subject to 
impurity scattering. 
 The Hamiltonian is given  by 
\begin{eqnarray} 
 H = \int d^3 x  
  \left[ \frac{\hbar^2}{2m} (\nabla c^\dagger)(\nabla c) 
  + c^\dagger V_{\rm imp}(x) \, c \, \right] 
  + H_{sd}  ,
\label{eq:H_total}
\end{eqnarray}
with 
\begin{eqnarray}
  V_{\rm imp} 
=  u_{\rm i} \sum_i \delta ({\bm r} - {\bm R}_i) 
 + u_{\rm s} \sum_j ({\bm S}_j \!\cdot\! {\bm \sigma}) \,
   \delta ({\bm r} - {\bm R}_j)  . 
\label{eq:V_imp} 
\end{eqnarray}
 We focus on the case that the magnetization is static, 
and basically uniform except for a small transverse deviation ${\bm u}$: 
\begin{eqnarray}
&{}& {\bm n}({\bm r}) 
= \hat z + {\bm u} ({\bm r}) 
= \hat z + {\bm u}_{\bm q} \, {\rm e}^{i {\bm q} \cdot {\bm r}} , 
\label{eq:u}
\end{eqnarray}
where ${\bm u} \perp \hat z$, $|{\bm u}| \ll 1$,  
and calculate the spin density to first order in ${\bm u}_{\bm q}$ and ${\bm q}$.  
 This is sufficient to determine the coefficients of each torque.\cite{TSBB06,KTS06,Duine07} 
 The impurity potential $V_{\rm imp}$ is treated in the Born approxination for the self-energy 
combined with ladder-type vertex corrections. 
 The renormalized Green function (for ${\bm u}={\bm 0}$) is given by 
\begin{eqnarray}
 G_{{\bm k} \sigma} (z) =   
  \left[ z + \mu - \hbar^2 {\bm k}^2/2m + M \sigma 
       + i \gamma_\sigma {\rm sgn}({\rm Im} z) \, \right]^{-1} ,
\label{eq:G}
\end{eqnarray}
with broadening (damping) 
\begin{eqnarray}
\gamma_\sigma = \frac{\hbar}{2\tau_\sigma} 
= \pi n_{\rm i} u_{\rm i}^2 \nu_\sigma 
+ \frac{\pi}{3} n_{\rm s} u_{\rm s}^2 S_{\rm imp}^2 (\nu_\sigma  + 2\nu_{\bar\sigma}) , 
\label{eq:gamma}
\end{eqnarray}
where $\tau_\sigma$ is the spin-dependent scattering lifetime, 
$n_{\rm i}$ and $u_{\rm i}$ ($n_{\rm s}$ and $u_{\rm s} S_{\rm imp}$) denote concentration and 
scattering potential of the normal (magnetic) impurities, respectively, and 
 $\nu_\sigma$ is the density of states of spin-$\sigma$ electrons. 
 At low enough temperatures, the chemical potential $\mu$ equals the Fermi energy $\varepsilon_{\rm F}$. 
We also define Fermi energies for each spin, $\sigma = \pm 1$, by 
$\varepsilon_{{\rm F}\sigma} = \varepsilon_{\rm F} + M \sigma$. 
 As in Ref.~\onlinecite{KTS06}, 
we assume a good ferromagnetic metal and retain 
only terms in the lowest nontrivial order in 
$\gamma_\sigma / (\mu + \sigma M)$ and 
$\gamma_\sigma / M$ 
 (which are collectively denoted by $\gamma$); 
explicitly, they are ${\cal O}(\gamma^{-1})$ for the spin-transfer torque and 
${\cal O}(\gamma^0)$ for the dissipative correction ($\beta$-term).

\section{Formulation}
\label{Formulation}

 Thermally-induced torques (induced by a temperature gradient, $\nabla T$) can be calculated 
analogously to ordinary current-induced torques (induced by an applied electric field ${\bm E}$). 
 In this section, we outline the formulation for both torques.

 Let us consider the general case in which conduction electrons in 
a ferromagnet are subject to an applied electric field (${\bm E}$), 
chemical-potential gradient (${\bm \nabla} \mu$), 
temperature gradient (${\bm \nabla} T$), 
and applied gravitational field ($-{\bm \nabla} \psi$). 
 The gravitational potential $\psi$ was introduced by Luttinger \cite{Luttinger64} 
as a field which couples to the local energy density, $h(x)$, 
thus driving an energy flow, ${\bm j}_E$. 
 Here we introduce it as a field coupling to $h(x) - \mu \, n (x)$, 
where $n(x)$ is the (local) number density, so that 
it drives heat current, ${\bm j}_Q = {\bm j}_E - \mu {\bm j}$, 
which is just a rearrangement to simplify the equations. 
 Then the {\it non-equilibrium part} of the transverse spin polarization, 
which is responsible for (non-equilibrium) spin torques, can be written as 
\begin{eqnarray}
 \langle \hat\sigma^\alpha_\perp ({\bm q}) \rangle_{\rm ne} 
= \chi_i^\alpha \left( E_i + \frac{1}{e} \nabla_i \mu \right) 
 + \chi_{Q,i}^\alpha \left( - \frac{\nabla_i T}{T} - \nabla_i \psi \right) , 
\nonumber \\
\label{eq:phenomenology}
\end{eqnarray}
where $\chi_i^\alpha$ and $\chi_{Q,\, i}^\alpha$ are linear-response coefficients 
with $\alpha$ and $i$ being spin and spatial indices, respectively. 
(In Eq.~(\ref{eq:phenomenology}), sum over $i=x,y,z$ is assumed.) 
 We use the same coefficient for $E_i$ and $\nabla_i \, \mu /e$, 
as well as for $\nabla_i T/T$ and $\nabla_i \, \psi$. 
 This can be justified by an argument {\it \`a la} Einstein:\cite{Einstein05,Luttinger64} 
under static, finite wavelength, and longitudinal perturbation, 
the system is in an equilibrium state, implying that torques {\it of non-equilibrium origin} should not arise.

 The coefficients of the mechanical perturbations 
($E_i$ and $-\nabla_i \, \psi$) are given by the standard 
Kubo formula \cite{Kubo57, Luttinger64} 
\begin{eqnarray}
 \chi_i^\alpha 
&=& \lim_{\omega \to 0}
  \frac{ K_i^\alpha ({\bm q},\omega +i0) - K_i^\alpha ({\bm q},0)}{i\omega} , 
\label{eq:chi}
\\
 \chi_{Q,i}^\alpha 
&=& \lim_{\omega \to 0}
  \frac{ K_{Q,i}^\alpha ({\bm q},\omega +i0) - K_{Q,i}^\alpha ({\bm q},0)}{i\omega} , 
\label{eq:chiQ}
\end{eqnarray}
where the response functions (see Appendix \ref{Linear Response}) are obtained from 
\begin{eqnarray}
 K_i^\alpha ({\bm q}, i\omega_\lambda ) 
&=& -e \int_0^\beta d\tau \, {\rm e}^{i\omega_\lambda \tau} \, 
    \langle \, {\rm T}_\tau \, \hat\sigma_\perp^\alpha ({\bm q}, \tau) 
            \, J_i \, \rangle ,  \ \ \ 
\label{eq:K_1}
\\
 K_{Q,\, i\,}^\alpha ({\bm q}, i\omega_\lambda ) 
&=& \int_0^\beta d\tau \, {\rm e}^{i\omega_\lambda \tau} \, 
    \langle \, {\rm T}_\tau \, \hat\sigma_\perp^\alpha ({\bm q}, \tau) 
            \, J_{Q,i} \, \rangle , 
\label{eq:KQ_1}
\end{eqnarray}
by the analytic continuation, $i\omega_\lambda \to \hbar\omega + i0$, 
where $\omega_\lambda = 2\pi \lambda k_{\rm B}T$ and $\beta = (k_{\rm B}T)^{-1}$.\cite{com:beta} 
 Here ${\bm J}$ is the total charge current (in units of $-e$) 
and ${\bm J}_Q$ is the total heat current; they are given 
by the volume integral of the corresponding current densities; 
\begin{eqnarray}
&& \hskip -7mm 
 {\bm j} (x) = \frac{\hbar}{2mi} 
 \lim_{x'\to x} (\nabla' -\nabla)  c^\dagger (x) \, c (x') , 
\label{eq:j}
\\ 
&& \hskip -7mm 
 {\bm j}_Q (x) = \frac{i \hbar}{4m} 
 \lim_{x'\to x} (\nabla' -\nabla)(\partial_{\tau'} - \partial_{\tau}) 
  c^\dagger (x) \, c (x') ,
\label{eq:jQ}
\end{eqnarray}
where $x=({\bm r},\tau)$ and $x'=({\bm r}',\tau')$. 
 Note that the expression (\ref{eq:jQ}) is written in imaginary time, $\tau$.

 The response functions, $ K_i^\alpha$ and $ K_{Q,\, i\,}^\alpha$, 
are non-zero in the presence of magnetization textures, 
Eq.~(\ref{eq:u}), and we extract $u^\beta$ and $q_j$ 
from $K_i^{\alpha}$ and $K_{Q,i}^{\alpha}$. 
 In the next section, we derive the forms 
\begin{eqnarray}
  K_i^{\alpha} ({\bm q}, i\omega_\lambda ) 
&=& - eM^{-1} \, 
  (\tilde b \, \delta^{\alpha\beta} + \tilde a \, \varepsilon^{\alpha\beta}) \, 
  \omega \, q_i u^\beta_{\bm q} , 
\label{eq:K_i_expansion}
\\
  K_{Q,i}^{\alpha} ({\bm q}, i\omega_\lambda ) 
&=& M^{-1}  \, 
  (\tilde b_T \, \delta^{\alpha\beta} + \tilde a_T \, \varepsilon^{\alpha\beta}) \, 
  \omega \, q_i u^\beta_{\bm q} , 
\label{eq:K_Qi_expansion}
\end{eqnarray}
where $\delta^{\alpha\beta}$ is the Kronecker's delta and 
$\varepsilon^{\alpha\beta}$ is the antisymmetric tensor 
(with $\varepsilon^{xy}=1$) in two dimensions, 
while $\tilde a, \tilde b, \tilde a_T^{\phantom{\dagger}}$ and 
$\tilde b_T^{\phantom{\dagger}}$ 
are yet unspecified coefficients. 
 These expressions indeed lead to the torques given by 
Eqs.~(\ref{eq:t_el_0}) and (\ref{eq:t_th_0}), with 
\begin{eqnarray}
&{}& {\bm v}_{\rm s} 
= - \frac{\tilde a }{s_{\rm tot}} \, ( e{\bm E} + \nabla \mu ), 
\ \ \ \ \ \   
\beta = \tilde b / \tilde a , 
\label{eq:v_s}
\\ 
&{}& {\bm v}_T 
= - \frac{\tilde a_T^{\phantom{\dagger}}}{s_{\rm tot}}  \, 
  \left( \frac{\nabla T}{T} + \nabla \psi \right) , 
\ \ \   
\beta_T^{\phantom{\dagger}} 
 = \tilde b_T^{\phantom{\dagger}} / \tilde a_T^{\phantom{\dagger}} . 
\label{eq:v_T}
\end{eqnarray}
 The calculation of the coefficients 
$\tilde a, \tilde b, \tilde a_T^{\phantom{\dagger}}$ and 
$\tilde b_T^{\phantom{\dagger}}$ 
in Eqs.~(\ref{eq:K_i_expansion}) and (\ref{eq:K_Qi_expansion}) is the subject of the next two sections.

 Before proceeding, we show that the two cases (electrical and thermal) can actually be calculated simultaneously. 
 In Eqs.~(\ref{eq:K_1}) and (\ref{eq:KQ_1}), 
the (imaginary-) time evolution and thermal average are determined by $H$. 
 Since this is a one-body Hamiltonian, $\dot c = [c, H ]/i\hbar$ is also a one-body operator. 
 We therefore can use Wick's theorem to factorize 
\begin{eqnarray}
&& \hskip -7mm 
 K_{i}^\alpha ({\bm q}, i\omega_\lambda ) 
=  e T\sum_n 
   \sum_{{\bm k}, {\bm k}'} v_{{\bm k}, \, i} \,  
   {\rm tr} \, [ \sigma^\alpha 
   {\cal G}_{{\bm k}'+{\bm q},  {\bm k}}^+ 
   {\cal G}_{{\bm k} , {\bm k}'} ] , 
\label{eq:K_2}
\end{eqnarray}
and 
\begin{eqnarray}
&& 
 K_{Q,\, i\,}^\alpha ({\bm q}, i\omega_\lambda ) 
\nonumber \\
&&=  -T\sum_n \left( i\varepsilon_n + i\omega_\lambda /2 \right) 
   \sum_{{\bm k}, {\bm k}'} v_{{\bm k}, \, i} \,  
   {\rm tr} \, [ \sigma^\alpha 
   {\cal G}_{{\bm k}'+{\bm q},  {\bm k}}^+ 
   {\cal G}_{{\bm k} , {\bm k}'} ]
\nonumber 
\\
&& \hskip 5mm 
 + \frac{1}{2} \, T\sum_n \sum_{\bm k} 
     {\rm tr} \, [ \sigma^\alpha 
   ( {\cal G}^+ \hat v_i 
   + \hat v_i  \, {\cal G} )_{{\bm k}+{\bm q}, {\bm k}} ] .
\label{eq:KQ_2}
\end{eqnarray}
 Here, 
${\cal G}_{{\bm k}\sigma , {\bm k}'\sigma'} 
\equiv {\cal G}_{{\bm k}\sigma , {\bm k}'\sigma'} (i\varepsilon_n) 
\equiv - \int_{\,0}^\beta d\tau \, {\rm e}^{i\varepsilon_n \tau} 
    \langle \, {\rm T} \, 
    c_{{\bm k}\sigma}^\dagger (\tau) \, 
    c_{{\bm k}'\sigma'}^{\phantom{\dagger}} \rangle $ 
is the exact Green function of $H$ (before the impurity average is taken), 
${\cal G}^+$ is the one with frequency $i\varepsilon_n + i\omega_\lambda$, 
${\bm v}_{\bm k} = \hbar {\bm k}/m$ is the electron velocity, 
and \lq\lq tr'' means trace in spin space. 
 (Since $H$ includes ${\bm u}({\bm r})$ and $V_{\rm imp}$, ${\cal G}$ has 
off-diagonal components in both spin and wavevector.)
 In deriving Eq.~(\ref{eq:KQ_2}), we used the relation, 
\begin{eqnarray}
  \langle \, {\rm T}_\tau \, c\, (\tau) \, \dot c^\dagger \, \rangle 
= -\langle \, {\rm T}_\tau \, \dot c\, (\tau) \, c^\dagger \, \rangle 
= \frac{d}{d\tau} {\cal G}(\tau) + \delta (\tau) . 
\end{eqnarray}
 Since the last term of Eq.~(\ref{eq:KQ_2}) 
does not depend on $\omega_\lambda$ 
(after summing over $\varepsilon_n$), 
it does not contribute to the result 
[see Eq.~(\ref{eq:chiQ})], and can be dropped. 
 Thus we are left only with the first term of Eq.~(\ref{eq:KQ_2}), 
showing that the heat-current vertex is simply governed by the factor 
$( i\varepsilon_n + i\omega_\lambda /2 ) \, {\bm v}_{\bm k}$. 
 We confirmed this statement starting from an explicit expression for the heat current  
(without using the time derivative) in Appendix \ref{Cancellation}.  
(For many-body Hamiltonians, see Ref.~\onlinecite{Kontani03}.)

\section{Explicit Calculation}
\label{Calculation}

\begin{figure}[t]
  \begin{center}
  \includegraphics[scale=0.32]{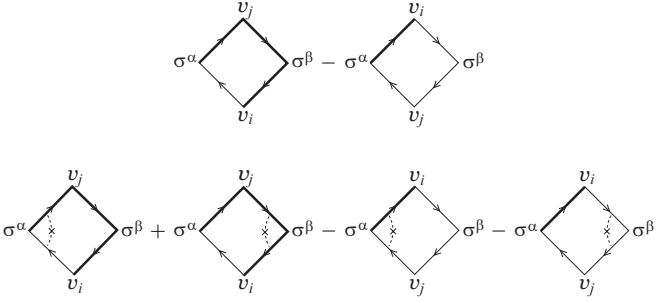}
  \vskip 0mm
  \end{center}
\caption{ 
   Diagrammatic expressions for the coefficient  
$K_{ij}^{\alpha\beta}$ and $K_{Q,ij}^{\alpha\beta}$ that govern the transverse spin polarization, 
$\langle \hat\sigma^\alpha_\perp ({\bm q}) \rangle_{\rm ne}$, 
which is linear in $E_i$ (or $-\nabla_i T /T$), $q_j$ and $u^\beta$, 
in the presence of current flow 
(induced by either electric field $E_i$ or temperature gradient 
 $\nabla_i T$) and magnetization texture ($q_j u^\beta$). 
 The velocity vertices $v_i$ and $v_j$ are associated with 
$E_i$ and $q_j$, respectively. 
 In the thermally-induced torque ($K_{Q,ij}^{\alpha\beta}$), 
the vertex $v_i$ is multiplied by $i(\varepsilon_n + \omega_\lambda /2)$.
 The thick (thin) solid lines represent electrons with Matsubara frequency 
$i\varepsilon_n + i\omega_\lambda$ ($i\varepsilon_n$).
 The dotted line with a cross represents scattering 
by non-magnetic or magnetic impurities.}
\label{fig:susceptibility_current_uni}
\end{figure}

 We calculate the torque coefficients, $K_{ij}^{\alpha\beta}$ and $K_{Q,ij}^{\alpha\beta}$ 
[Eqs.~(\ref{eq:K_i_expansion}) and (\ref{eq:K_Qi_expansion}))] 
 by first extracting $q_j$ and $u^\beta_{\bm q}$ from $K_i^{\alpha}$ and $K_{Q,i}^{\alpha}$ as 
\begin{eqnarray}
&{}&  K_i^{\alpha} ({\bm q}, i\omega_\lambda ) 
= - eM K_{ij}^{\alpha\beta} (i\omega_\lambda ) \, q_j u^\beta_{\bm q} , 
\\
&{}&  K_{Q,i}^{\alpha} ({\bm q}, i\omega_\lambda ) 
= M K_{Q,ij}^{\alpha\beta} (i\omega_\lambda ) \, q_j u^\beta_{\bm q} ,
\end{eqnarray}
where summing over $j=x,y,z$ and $\beta=x,y$ is implied.
 To the lowest nontrivial order in $\gamma$, 
they are expressed diagrammatically in Fig.~1, and read 
\begin{eqnarray}
&& \hskip -7mm 
 K_{ij}^{\alpha\beta} (i\omega_\lambda ) 
= T\sum_n 
 \varphi_{ij}^{\alpha\beta} (i\varepsilon_n+i\omega_\lambda, i\varepsilon_n) ,
\label{eq:K_ij_0}
\\
&& \hskip -7mm 
 K_{Q,ij}^{\alpha\beta} (i\omega_\lambda ) 
= T\sum_n (i\varepsilon_n+i\omega_\lambda/2) \, 
 \varphi_{ij}^{\alpha\beta} (i\varepsilon_n+i\omega_\lambda, i\varepsilon_n) , \ \ \  
\label{eq:K_Qij_0}
\end{eqnarray}
where \cite{com4} 
\begin{eqnarray}
&& 
 \varphi_{ij}^{\alpha\beta} (i\varepsilon_n+i\omega_\lambda, i\varepsilon_n) 
\nonumber \\ 
&& 
=  \sum_{{\bm k}} v_i v_j 
  \biggl\{ 
   {\rm tr} \left[ \sigma^\alpha G^+ G^+ \sigma^\beta G^+ G \right] 
 - {\rm tr} \left[ \sigma^\alpha G^+ G \sigma^\beta G G \right] 
  \biggr\} 
\nonumber \\ 
&& 
+ \, \tilde \Gamma_0 \sum_{{\bm k},{\bm k}'} v_i v_j 
  \biggl\{ 
   {\rm tr} \left[ (G' \sigma^\alpha G^{+\prime}) G^+ G^+ \sigma^\beta G^+ G \right] 
\nonumber \\ 
&& \hskip 22mm 
-   {\rm tr} \left[ (G' \sigma^\alpha G^{+\prime}) G^+ G \sigma^\beta G G  \right] 
\nonumber \\ 
&& \hskip 22mm 
+  {\rm tr} \left[ \sigma^\alpha G^+ G^+ 
                  (G^{+\prime} \sigma^\beta G^{+\prime}) \, G^+ G \right] 
\nonumber \\ 
&& \hskip 22mm 
 - {\rm tr} \left[ \sigma^\alpha G^+ G (G' \sigma^\beta G') \, G G \right] 
  \biggr\} . 
\label{eq:varphi_0}
\end{eqnarray}
 Here, the following notation has been used: 
$G^+ = G_{\bm k}(i\varepsilon_n + i\omega_\lambda )$, 
$G = G_{\bm k}(i\varepsilon_n )$, 
$G^{+\prime} = G_{{\bm k}'}(i\varepsilon_n + i\omega_\lambda )$, 
$G' = G_{{\bm k}'}(i\varepsilon_n )$, 
$v_i = \hbar k_i/m$, and 
$\tilde \Gamma_0 = n_{\rm i} u_{\rm i}^2 - n_{\rm s} u_{\rm s}^2 S_{\rm imp}^2/3$. 
 The electrically-induced torques, 
Eqs.~(\ref{eq:K_ij_0}) and (\ref{eq:varphi_0}), 
have been studied in Ref.~\onlinecite{KTS06}. 
 New in this paper is introduction and treatment of Eq.~(\ref{eq:K_Qij_0}).

 After the analytic continuation,
$i\omega_\lambda \to \hbar\omega + i0$, 
we expand $K_{ij}^{\alpha\beta}$ and $K_{Q,ij}^{\alpha\beta}$ with respect to $\omega$ as 
\begin{eqnarray}
&& \hskip -7mm 
 K(\omega + i0 ) - K(0) 
\nonumber \\
&& \hskip -7mm 
= \frac{i \hbar \omega}{2\pi} 
\int_{-\infty}^\infty d\varepsilon 
  \left( - \frac{\partial f}{\partial \varepsilon} \right) 
  \left\{ {\rm Re} [\varphi^{(1)} (\varepsilon,\varepsilon)] 
         - \varphi^{(2)} (\varepsilon,\varepsilon) 
   \right\} 
\nonumber \\
&& \hskip -3mm 
 - \frac{\hbar \omega}{2\pi} 
   \int_{-\infty}^\infty d\varepsilon \, f (\varepsilon) 
     \left. 
\left( \partial_\varepsilon - \partial_{\varepsilon'} \right) 
   {\rm Im} \bigl[ \varphi^{(1)} (\varepsilon, \varepsilon') \bigr] 
   \right|_{\varepsilon' = \varepsilon} 
\nonumber \\
&& \hskip -3mm 
   + {\cal O}(\omega^2)
, 
\label{eq:K_ij_w-linear}
\end{eqnarray}
where $f(\varepsilon)$ is the Fermi-Dirac distribution function, and 
$\partial_\varepsilon = \partial/\partial \varepsilon$, 
$\partial_{\varepsilon'} = \partial/\partial \varepsilon'$. 
 In Eq.~(\ref{eq:K_ij_w-linear}), 
$\varphi = \varphi_{ij}^{\alpha\beta}(\varepsilon,\varepsilon')$ 
for $K=K_{ij}^{\alpha\beta}$, 
and $\varphi = [(\varepsilon + \varepsilon')/2] \, 
 \varphi_{ij}^{\alpha\beta}(\varepsilon,\varepsilon')$ 
for $K=K_{Q,ij}^{\alpha\beta}$; 
the superscripts $(1), (2)$ and $(3)$ on $\varphi$ express 
the analytic branches continued as 
$G^+G \to G^{\, {\rm R}} G^{\, {\rm R}}$, 
$G^{\, {\rm R}} G^{\, {\rm A}}$ and $G^{\, {\rm A}} G^{\, {\rm A}}$, 
respectively.

 After some manipulations, the coefficients adopt the form, 
\begin{eqnarray}
&& \hskip -7mm 
 \tilde a 
= \int_{-\infty}^\infty d\varepsilon 
   \left( - \frac{\partial f}{\partial \varepsilon} \right) A(\varepsilon) , 
\label{eq:a}
\\ 
&& \hskip -7mm 
 \tilde b 
= \int_{-\infty}^\infty d\varepsilon 
   \left( - \frac{\partial f}{\partial \varepsilon} \right) B(\varepsilon) 
- \int_{-\infty}^\infty d\varepsilon 
   f(\varepsilon ) \, \partial_\varepsilon C(\varepsilon) ,
\label{eq:b}
\end{eqnarray}
for electrically-induced torques, and 
\begin{eqnarray}
&& \hskip -7mm 
 \tilde a_T 
= \int_{-\infty}^\infty d\varepsilon 
   \left( - \frac{\partial f}{\partial \varepsilon} \right) 
   \varepsilon A(\varepsilon) , 
\label{eq:aT}
\\ 
&& \hskip -7mm 
 \tilde b_T 
= \int_{-\infty}^\infty d\varepsilon 
   \left( - \frac{\partial f}{\partial \varepsilon} \right) 
  \varepsilon B(\varepsilon) 
- \int_{-\infty}^\infty d\varepsilon 
   f(\varepsilon ) \, \varepsilon \, \partial_\varepsilon C(\varepsilon) , 
\label{eq:bT}
\end{eqnarray}
for thermally-induced torques. 
 The terms containing $( - \partial f/\partial \varepsilon )$ are called 
\lq\lq Fermi-surface terms'', and those with $f(\varepsilon ) $ as  \lq\lq Fermi-sea terms''. \cite{com_sea}
 This separation is not unique in a strict sense, but convenient in practice 
(at least in the present context) if defined symmetrically ($\varepsilon \pm \omega /2$) 
as in Eq.~(\ref{eq:K_ij_w-linear}). 
 The functions $A, B$ and $C$ are given by 
\begin{eqnarray}
&& \hskip -7mm 
 A (\varepsilon) 
= \frac{M^2}{\pi} \sum_\sigma \sigma {\rm Re} L_\sigma (\varepsilon), 
\\ 
&& \hskip -7mm 
 B (\varepsilon) 
= \frac{M^2}{\pi} \sum_\sigma {\rm Im} L_\sigma (\varepsilon), 
\\ 
&& \hskip -7mm 
 C (\varepsilon) 
=  \frac{M^2}{\pi} {\rm Im} \sum_{{\bm k}} v_i v_j 
      \left( G_{{\bm k} \uparrow}^{\, {\rm R}} 
             G_{{\bm k} \downarrow}^{\, {\rm R}}  \right)^2 , 
\label{eq:C}
\end{eqnarray}
with 
\begin{eqnarray}
&& \hskip -7mm 
 L_\sigma (\varepsilon) 
=  \sum_{{\bm k}} v_i v_j 
            G_{{\bm k} \sigma}^{\, {\rm R}} 
           (G_{{\bm k} \bar\sigma}^{\, {\rm R}})^2 
            G_{{\bm k} \sigma}^{\, {\rm A}} 
\nonumber \\
&& \ \ \ \ 
  \times  \left\{ 1 + \tilde \Gamma_0 
               \sum_{{\bm k}'} 
               G_{{\bm k}' \bar\sigma}^{\, {\rm R}} 
              (G_{{\bm k}' \sigma}^{\, {\rm R}}
             + G_{{\bm k}' \sigma}^{\, {\rm A}} ) 
         \right\} .
\label{eq:L}
\end{eqnarray}
 In Eqs.~(\ref{eq:C}) and (\ref{eq:L}), all Green functions share the frequency argument $\varepsilon$. 
 Equations (\ref{eq:a})-(\ref{eq:bT}) can be rewritten as 
\begin{eqnarray}
&& \hskip -7mm 
 \tilde a = A_0  , \ \ \ \ \ 
 \tilde b = B_0 - C_0  , 
\label{eq:ab}
\\ 
&& \hskip -7mm 
 \tilde a_T = A_1  , \ \ \ \ \ 
 \tilde b_T = B_1 - C_1 + c  , 
\label{eq:abT}
\end{eqnarray}
where 
\begin{eqnarray}
&& \hskip -7mm 
 A_n  
=  \int_{-\infty}^\infty d\varepsilon 
   \left( - \frac{\partial f}{\partial \varepsilon} \right) 
   \varepsilon^n A(\varepsilon) , 
\label{eq:An}
\end{eqnarray}
and similarly for $B_n$ and $C_n$, with
\begin{eqnarray}
 c = \int_{-\infty}^\infty d\varepsilon 
   f(\varepsilon ) \, C(\varepsilon)  . 
\label{eq:c}
\end{eqnarray} 
 Using Eq.~(\ref{eq:t_sd_0}), the torques are obtained as 
\begin{eqnarray}
&& \hskip -10mm 
 {\bm t}_{\rm el} 
= [ A_0 \, \partial_i {\bm n} + (B_0-C_0) ({\bm n} \times \partial_i {\bm n}) ] \, eE_i ,
\label{eq:t_phi}
\\
&& \hskip -10mm 
 {\bm t}^{(\psi)} 
= [ A_1 \, \partial_i {\bm n} + (B_1-C_1+c) ({\bm n} \times \partial_i {\bm n}) ] \, \partial_i \psi .
\label{eq:t_psi}
\end{eqnarray}
 Note that as $T \to 0$, $A_1, B_1$ and $C_1$ vanish, but $c$ remains finite.

 The $c$-term in Eq.~(\ref{eq:t_psi}),  
\begin{eqnarray}
 \Delta {\bm t}^{(\psi)}  
\equiv c \, ({\bm n} \times \partial_i {\bm n}) \, \partial_i \psi , 
\label{eq:Dt_psi_unwanted}
\label{eq:unwanted}
\end{eqnarray}
is problematic for the following reasons. 
 If the Einstein-Luttinger relation (\ref{eq:phenomenology}) is applied, it leads to a thermally-induced torque  
\begin{eqnarray}
 \Delta {\bm t}_{\rm th} 
= c \, ({\bm n} \times \partial_i {\bm n}) \, \partial_i T /T , 
\end{eqnarray}
which diverges as $T \to 0$ (since $c$ is finite as $T \to 0$).  
 This contradicts the thermodynamic law (Nernst theorem) that thermally-induced effects should vanish with temperature.  
 Also, the predicted finite $\beta_T$ even in the absence of spin relaxation  violates the spin conservation. 
 Therefore, Eq.~(\ref{eq:unwanted}) must be carefully reconsidered.

\section{Subtraction of Equilibrium Components}
\label{Eq}

 To settle the problem encountered in the last section, 
we note that the combination $-\nabla \psi - \nabla T/T$ in Eq.~(\ref{eq:phenomenology}) 
should be applied only to {\it non-equilibrium} components that must be identified beforehand.

 Even at equilibrium, {\it i.e.}, without external fields $E_i=0$ and $\psi = 0$, 
a finite spin density 
$\langle \hat {\bm \sigma} \rangle_{\rm eq} = (c/M) \, \nabla^2 {\bm n}$ exists, which 
corresponds to the exchange-stiffness torque 
\begin{eqnarray}
 {\bm t}_{\rm eq} = c \, ({\bm n} \times \nabla^2 {\bm n}) . 
\label{eq:t_eq}
\end{eqnarray}
 The coefficient $c$ is the same as in Eq.~(\ref{eq:c}), and represents the contribution of the conduction electrons
to the exchange-stiffness constant; see Appendix \ref{Exchange} for the calculation.

 This equilibrium torque is modified by $\psi$ in two ways. 
 First, the torque formula, Eq.~(\ref{eq:t_sd_0}), acquires an additional factor 
\begin{eqnarray}
 {\bm t}_{sd}^{(\psi)} = M {\bm n}(x) \times \langle \hat {\bm \sigma} (x) \rangle (1+\psi) , 
\label{eq:t_sd_1}
\end{eqnarray}
since the $s$-$d$ coupling, and hence the effective field $\sim \delta H_{sd}/ \delta {\bm n}$, is multiplied by $1+\psi$. 
 Secondly, the spin density $\langle \hat {\bm \sigma} (x) \rangle$ may be modified by $\psi$ 
(on top of a term proportional to $\partial_i \psi$). 
 It turns out, however, that no such terms arise in $\langle \hat {\bm \sigma} (x) \rangle$; 
see Eq.~(\ref{eq_app:s_psi_1}) for an explicit expression, 
and Appendix A for a formal derivation.  
 From a general point of view, this is owed to the adiabatic nature of the Kubo formula 
and the conserved nature of the perturbed quantity (energy). 
This is shown in Appendix \ref{Thermodynamics}. 
 Therefore, the equilibrium spin density $\langle \hat {\bm \sigma} \rangle_{\rm eq}$ 
in the previous paragraph (for $\psi=0$) is not modified by a uniform $\psi$ 
(namely, in the zeroth-order gradient of $\psi$). 
Therefore, using Eq.~(\ref{eq:t_eq}) in Eq.~(\ref{eq:t_sd_1}), we obtain  
\begin{eqnarray}
 {\bm t}_{{\rm eq'}}^{(\psi)} = c \, ({\bm n} \times \nabla^2 {\bm n})(1+\psi) .
\label{eq:t_sd_psi}
\end{eqnarray}
 (The suffix eq$'$ means that this term does not exhaust the equilibrium torque.) 
 The total equilibrium torque is the sum of Eq.~(\ref{eq:t_sd_psi}) and Eq.~(\ref{eq:t_psi}); 
the former contains all torques proportional to $\psi$, 
and the latter those proportional to $\partial_i \psi$. 
 Focussing on terms containing $c$ 
\begin{eqnarray}
 {\bm t}_{{\rm eq'}}^{(\psi)} + \Delta {\bm t}^{(\psi)}  
=  - \partial_i \, {\bm j}_{{\rm s}, i}^{(\psi)} , 
\label{eq:scenario_1}
\end{eqnarray}
where 
\begin{eqnarray}
  {\bm j}_{{\rm s}, i}^{(\psi)} = - c \, ({\bm n} \times \partial_i {\bm n}) (1+\psi )
\label{eq:js}
\end{eqnarray}
is the spin-current density (in the classical magnetization texture formed by localized spin system) in the presence of $\psi$. 
 The right-hand side of Eq.~(\ref{eq:scenario_1}) represents the (generalized) exchange-stiffness torque in the presence of $\psi$, 
{\it which we identify as the total equilibrium torque}. 
 By subtracting this equilibrium component,  we identify 
 the non-equilibrium component to be Eq.~(\ref{eq:t_psi}) without the offensive $c$-term. 
 The replacement, $\partial_i \psi \to \partial_i T/T$, should be enforced only in this non-equilibrium component 
 such that 
\begin{eqnarray}
 {\bm t}_{\rm th}  
= [ A_1 \, \partial_i {\bm n} + (B_1-C_1) ({\bm n} \times \partial_i {\bm n}) ] \, \partial_i T/T ,
\label{eq:t_th_2}
\end{eqnarray}
behaves regularly (namely, vanishes) as $T \to 0$.

 The above procedure, Eqs.~(\ref{eq:t_sd_psi})-(\ref{eq:t_th_2}), 
may be better understood by subjecting an insulating ferromagnet (without mobile $s$ electrons) to $\psi$. 
 Its Lagrangian is given by 
\begin{eqnarray}
 L = \int d^3 x \left\{ \hbar S \dot \varphi \cos\theta - \frac{J}{2} (\nabla {\bm n})^2 (1+\psi) \right\} ,  
\label{eq:L_ins}
\end{eqnarray}
where $(\theta,\varphi)$ represents the direction of ${\bm n}$. 
 Note that $\psi$ couples only to the energy density $J (\nabla {\bm n})^2/2$ 
(anisotropy, damping, etc.~are neglected for simplicity), and not to the kinetic term (first term).   
 The variational principle leads to the equation of motion\cite{TKS08}  
\begin{eqnarray}
  \hbar S \dot {\bm n} 
= J \, \partial_i[ ({\bm n} \times \partial_i {\bm n})(1+\psi)] ,
\label{eq:LLG3}
\end{eqnarray}
whose right-hand side precisely corresponds to Eq.~(\ref{eq:scenario_1}).   
 This supports the identification of the equilibrium torque in the preceding paragraph.

 A similar difficulty has been noted for thermal transport in magnetic fields.   
 To resolve it, the authors of Refs.~\onlinecite{Oji85}-\onlinecite{Qin11} proposed 
to extract the transport current by subtracting the magnetization current, 
and then to apply the substitution $\partial_i \psi \to \partial_i T/T$ to the transport current. 
 In this procedure, it is essential that the expressions for electric and heat currents are modified 
by $\psi$ (as in Eq.~(\ref{eq:t_sd_1})).

\section{Result}
\label{Result}

 We thus arrive at expressions for the non-equilibrium torque 
${\bm t}_{\rm tot} = {\bm t}_{\rm el} + {\bm t}_{\rm th}$,  
\begin{eqnarray}
&& \hskip -10mm 
 {\bm t}_{\rm el} 
= [ A_0 \, \partial_i {\bm n} + (B_0-C_0) ({\bm n} \times \partial_i {\bm n}) ] \, eE_i ,
\label{eq:t_phi_1}
\\
&& \hskip -10mm 
 {\bm t}_{\rm th}  
= [ A_1 \, \partial_i {\bm n} + (B_1-C_1) ({\bm n} \times \partial_i {\bm n}) ] \, \partial_i T/T ,
\label{eq:t_psi_1}
\end{eqnarray}
where the coefficients are given by (\ref{eq:An}) with\cite{KTS06}
\begin{eqnarray}
&& \hskip -7mm 
 A(\varepsilon) = \frac{\hbar}{2e} \sigma_{\rm s}(\varepsilon) , \ \ \ 
 B(\varepsilon) - C(\varepsilon) = \beta (\varepsilon) \, \frac{\hbar}{2e} \sigma_{\rm s} (\varepsilon) , 
\end{eqnarray}
and thus
\begin{eqnarray}
&& \hskip -10mm 
 {\bm t}_{\rm el} 
= \frac{\hbar}{2} E_i 
 \int d\varepsilon \left( - \frac{\partial f}{\partial \varepsilon} \right)
 \sigma_{\rm s} (\varepsilon) [ \, \partial_i {\bm n} + \beta (\varepsilon) ({\bm n} \times \partial_i {\bm n}) ]  ,
\label{eq:t_phi_1}
\\
&& \hskip -10mm 
 {\bm t}_{\rm th}  
= \frac{\hbar}{2e} \frac{\nabla_i T}{T} 
  \int d\varepsilon \left( - \frac{\partial f}{\partial \varepsilon} \right) \varepsilon \, 
  \sigma_{\rm s} (\varepsilon) 
 [ \, \partial_i {\bm n} + \beta (\varepsilon) ({\bm n} \times \partial_i {\bm n}) ]  . 
\nonumber \\
\label{eq:t_psi_1}
\end{eqnarray}
 Here, $\sigma_{\rm s} (\varepsilon)$ is the \lq\lq spin conductivity'' and $\beta (\varepsilon )$ is dissipative correction, 
\begin{eqnarray}
 \sigma_{\rm s}  &=& \frac{e^2}{m} (n_\uparrow \tau_\uparrow - n_\downarrow \tau_\downarrow ) , 
\label{eq:sigma_s}
\\
 \beta &=& \frac{2\pi }{3} n_{\rm s} u_{\rm s}^2 S_{\rm imp}^2 \frac{\nu_\uparrow +\nu_\downarrow}{M} , 
\label{eq:beta}
\end{eqnarray}
evaluated at energy $\mu + \varepsilon$ 
(or $\varepsilon_{\rm F} + \varepsilon$ at low enough temperatures), 
with $n_\sigma$ being the density of spin-$\sigma$ electrons. 
 The relation between ${\bm t}_{\rm th}$ and ${\bm t}_{\rm el}$ may be symbolically written as  
\begin{eqnarray}
 {\bm t}_{\rm th} = \int d\varepsilon \left( - \frac{\partial f}{\partial \varepsilon} \right) \varepsilon \, 
 {\bm t}_{\rm el} (\varepsilon ) \, \Bigr|_{\, e{\bm E} \to \nabla T/T} ,
\label{eq:t_th_result}
\end{eqnarray}
where the electric field ${\bm E}$ in ${\bm t}_{\rm el}$ 
is replaced by the temperature gradient $\nabla T$ in ${\bm t}_{\rm th}$. 
 (${\bm t}_{\rm el}(\varepsilon )$ is defined by the total integrand of Eq.~(\ref{eq:t_phi_1}).) 

 For sufficiently low temperatures, the Sommerfeld expansion 
\begin{eqnarray}
 \int_{-\infty}^\infty d\varepsilon \, 
    F(\varepsilon) \left( - \frac{\partial f}{\partial \varepsilon} \right) 
= F(0) + \frac{\pi^2}{6} F''(0) (k_{\rm B}T)^2 + \cdots , 
\end{eqnarray}
can be used to evaluate as $A_0=A(0)$, $A_1 = (\pi^2/3) A'(0) (k_{\rm B}T)^2 $, etc.  
 Here, the prime originally refers to the $\varepsilon$-derivative, but it can be 
redefined to be the $\varepsilon_{\rm F}$-derivative, since $\varepsilon$ and $\varepsilon_{\rm F}$ 
appear only as $\varepsilon + \varepsilon_{\rm F}$ in the unperturbed Green function, Eq.~(\ref{eq:G}), 
and the factor $\varepsilon$ in 
Eqs.~(\ref{eq:aT}) and (\ref{eq:bT}) does not appear in $F''(0)$. 
 Hence 
\begin{eqnarray}
 \tilde a_T 
= \frac{\pi^2}{3} \frac{d \tilde a}{d\varepsilon_{\rm F}} (k_{\rm B}T)^2 , 
\ \ \ \ \ 
 \tilde b_T 
= \frac{\pi^2}{3} \frac{d \tilde b}{d\varepsilon_{\rm F}} (k_{\rm B}T)^2 ,  
\label{eq:ab_T_2}
\end{eqnarray}
or 
\begin{eqnarray}
 {\bm t}_{\rm th} = \frac{\pi^2}{3} (k_{\rm B}T)^2 
 \left. \frac{d}{d\varepsilon_{\rm F}} {\bm t}_{\rm el} \, \right|_{\, e{\bm E} \to \nabla T/T} . 
\label{eq:t_T_2}
\end{eqnarray}
 These are \lq Mott formulae' for the thermally-induced spin torque 
in terms of the $\varepsilon_{\rm F}$-derivative of the electrical counterpart.

 Explicitly, the total torque is written as
\begin{eqnarray}
\hskip -7mm 
 \tilde {\bm t}_{\rm tot} 
&=& \frac{\hbar}{2e s_{\rm tot}} \bigl\{ 
  ({\bm j}_{\rm s}^{\,{\rm tot}} \!\cdot\! {\bm \nabla})\,  {\bm n} 
  + \beta \, {\bm n} \times ({\bm j}_{\rm s}^{\,{\rm tot}} \cdot\! {\bm \nabla}) 
    \, {\bm n} 
\nonumber \\
&& \hskip 3cm 
  + \beta'  {\bm n} \times ( {\bm j}_{Q,{\rm s}} \cdot\! {\bm \nabla}) 
    \, {\bm n} 
 \bigr\}   , 
\label{eq:result}
\end{eqnarray}
where 
\begin{eqnarray}
&& \hskip -7mm 
  {\bm j}_{\rm s}^{\rm tot} = \sigma_{\rm s} \, ({\bm E} + {\cal S}_{\rm s} \nabla T ) , 
\ \ \ \ \   {\cal S}_{\rm s} 
= \frac{\pi^2 k_{\rm B}^2 }{3e} \,  \frac{\sigma_{\rm s}'}{\sigma_{\rm s}} \, T , 
\end{eqnarray}
with ${\cal S}_{\rm s}$ reflecting the spin dependence of the Seebeck coefficient, and 
\begin{eqnarray}
   {\bm j}_{Q,{\rm s}} 
= \frac{\pi^2 k_{\rm B}^2 }{3e} \, \sigma_{\rm s} \, T \, \nabla T 
\label{eq:jQs}
\end{eqnarray}
is the \lq spin-heat current', {\it i.e.}, spin-polarized part of the heat current (multiplied by $-e$).  
 The second and the third terms in the brackets of Eq.~(\ref{eq:result}) follow from 
$(\beta \sigma_{\rm s})' = \beta \sigma_{\rm s}' + \beta' \sigma_{\rm s}$. 
 While the first and the second terms are the ordinary spin-transfer torque and the $\beta$-term due to thermoelectric spin current, 
the third term (with $\beta'$) is the spin torque directly driven by the heat current.

\section{Application}
\label{Application}

 To illustrate the implications of the microscopic result, 
we consider now a temperature gradient without external electric field, ${\bm E}_{\rm ext}={\bm 0}$. 
 The spin torque depends on the type of the circuit (closed or open) 
because of the internal field ${\bm E}_{\rm int}$,
where ${\bm E} = {\bm E}_{\rm ext} + {\bm E}_{\rm int}$.\cite{Bauer09}  
 Total spin torque (\ref{eq:result}) may then be rewritten as
\begin{eqnarray}
 \tilde {\bm t}_{\rm tot} ({{\bm E}_{\rm ext}={\bm 0}}) 
= \frac{\hbar}{2e s_{\rm tot}} 
 (1 + \beta_T \, {\bm n} \times ) ({\bm j}_{\rm s}^T \cdot\! {\bm \nabla}) \, {\bm n} ,
\label{eq:result_2}
\end{eqnarray}
where ${\bm j}_{\rm s}^T$ is proportional to $\nabla T$ and $\beta_T$ is an effective beta parameter. 
 For a closed circuit, ${\bm E}={\bm 0}$, 
the thermal spin-transfer torque is governed by the thermoelectric spin current 
${\bm j}_{\rm s}^{T,{\rm closed}} = \sigma_{\rm s} {\cal S}_{\rm s} \nabla T$, 
and the thermal $\beta$-term by 
\begin{eqnarray}
  \beta_T^{\rm closed}  {\bm j}_{\rm s}^{T,{\rm closed}}  
&=& \frac{\pi^2 k_{\rm B}^2 }{3e} \, \beta \sigma_{\rm s} \left( \frac{\sigma_{\rm s}'}{\sigma_{\rm s}} +\frac{\beta'}{\beta}  \right) T \nabla T , 
\label{eq:closed}
\end{eqnarray}
where 
\begin{eqnarray}
\beta_T^{\rm closed} = \beta + \beta' \frac{\sigma_{\rm s}}{\sigma_{\rm s}'} . 
\end{eqnarray}
 For open circuits ${\bm j} = \sigma_{\rm c} ({\bm E} + {\cal S}_{\rm c} \nabla T) = {\bm 0}$ 
with $\sigma_{\rm c} = (e^2/m) (n_\uparrow \tau_\uparrow + n_\downarrow \tau_\downarrow )$ 
and ${\cal S}_{\rm c} = (\pi^2 k_{\rm B}^2 /3e) (\sigma_{\rm c}'/\sigma_{\rm c}) T $, 
the thermal spin-transfer torque is governed by 
${\bm j}_{\rm s}^{T,{\rm open}}  = \sigma_{\rm s} ({\cal S}_{\rm s} - {\cal S}_{\rm c}) \nabla T$. 
 The thermal $\beta$-term then reads
\begin{eqnarray}
   \beta_T^{\rm open}  {\bm j}_{\rm s}^{T,{\rm open}} 
= \frac{\pi^2 k_{\rm B}^2 }{3e} \, \beta \sigma_{\rm s} \left( - \frac{\sigma_{\rm c}'}{\sigma_{\rm c}} + \frac{\sigma_{\rm s}'}{\sigma_{\rm s}} +\frac{\beta'}{\beta}  \right) T \nabla T ,
\label{eq:open}
\end{eqnarray}
where 
\begin{eqnarray}
\beta_T^{\rm open} = \beta + \beta' \left( \frac{\sigma_{\rm s}'}{\sigma_{\rm s}} - \frac{\sigma_{\rm c}'}{\sigma_{\rm c}}  \right)^{-1} . 
\end{eqnarray}
 Thus, the thermal $\beta_T$ differs from the electrical one ($\beta$) when $\beta' \ne 0$.

 In the present model (\ref{eq:H_total}) with parabolic electron dispersion and high electron densities, $\sigma_{\rm s}$ depends on $\epsilon_{\rm F}$ only weakly\cite{com:sigma_s} and the thermoelectric spin current ($\propto \sigma_{\rm s}'$) is vanishingly small, whereas $\sigma_{\rm c}'/\sigma_{\rm c} \sim 1/\varepsilon_{\rm F}$ 
and  $\beta'/\beta = (\nu_\uparrow' + \nu_\downarrow')/(\nu_\uparrow + \nu_\downarrow) \sim 1/2\varepsilon_{\rm F}$ (if $\varepsilon_{\rm F} \pm M$ are not too small compared to $\varepsilon_{\rm F}$). 
 Therefore, in closed circuits, the thermal spin-transfer torque is dominated by the thermal $\beta$-term $\propto \beta' \sigma_{\rm s} T \nabla T$ driven by the spin-heat current, Eq.~(\ref{eq:jQs}).  
 By opening the circuits, both torques change sign by the effect of 
${\bm E}_{\rm int}$ $(\propto -\sigma_{\rm c}'/\sigma_{\rm c})$. 
 A domain wall is thus driven in mutually opposite directions in closed and open circuits. 
 In real materials, such features of course depend on the details of band structure.

\section{ General Aspects}
\label{Generalization}

 In this section, we draw some general conclusion out of the analysis in the previous sections. 
 For this purpose, it is convenient to shift the (off-shell) energy variable $\varepsilon$ as $\varepsilon \to \varepsilon - \mu$, 
so that the Fermi-Dirac distribution function is explicitly $\mu$-dependent but the Green functions are not. 
 Without introducing new functions, we redefine 
$f(\varepsilon) = ({\rm e}^{\beta (\varepsilon - \mu)}+1)^{-1}$ instead of $f(\varepsilon) = ({\rm e}^{\beta \varepsilon}+1)^{-1}$, 
and  $G(\varepsilon) = (\varepsilon - \varepsilon_{\bm k} + \cdots)^{-1}$ instead of $G(\varepsilon) = (\varepsilon + \mu - \varepsilon_{\bm k} + \cdots)^{-1}$, 
and similarly for $B(\varepsilon)$ and $C(\varepsilon)$. 
(We focus on $\tilde b$ and $\tilde b_T$.)

 Following Luttinger's prescription, we considered the linear response to a field $\psi$ which couples to the energy (or heat) density. 
 Thermal response functions have been obtained from the electrical response functions by simply introducing an  $(\varepsilon  -\mu)$-factor 
inside the $\varepsilon$-integral. 
 This \lq\lq $(\varepsilon  -\mu)$-factor prescription'' works well for the Fermi-surface term, 
\begin{eqnarray}
 \chi_{\rm el}^{\rm surface} 
&=& \int_{-\infty}^\infty d\varepsilon \left( - \frac{\partial f}{\partial \varepsilon} \right) B(\varepsilon) ,
\label{eq:chi_el_surf_1}
\\
 \chi_{\rm th}^{\rm surface} 
&=& \int_{-\infty}^\infty d\varepsilon\left( - \frac{\partial f}{\partial \varepsilon} \right) (\varepsilon - \mu) B(\varepsilon) . 
\label{eq:chi_th_surf_1}
\end{eqnarray}
 On the other hand, for the Fermi-sea terms, it leads to an unphysical contribution that can be repaired by
subtracting the equilibrium components, leading to
\begin{eqnarray}
 \chi_{\rm el}^{\rm sea} 
&=& \int_{-\infty}^\infty d\varepsilon f(\varepsilon ) D(\varepsilon) ,
\label{eq:chi_el_general}
\\
 \chi_{\rm th}^{\rm sea} 
&=& \int_{-\infty}^\infty d\varepsilon 
   f(\varepsilon ) \, (\varepsilon - \mu) D(\varepsilon) 
 - \int_{-\infty}^\infty d\varepsilon f(\varepsilon ) \, C(\varepsilon) , 
\nonumber \\
\label{eq:chi_th_general}
\end{eqnarray}
where $D(\varepsilon) \equiv -\partial_\varepsilon C(\varepsilon)$. 
 The first term in $\chi_{\rm th}^{\rm sea}$ includes the $(\varepsilon - \mu)$-factor  
for the heat (or heat-current) vertex, while the second term subtracts the equilibrium component. 
 By partial integration, 
\begin{eqnarray}
 \chi_{\rm th}^{\rm sea} 
&=& \int_{-\infty}^\infty d\varepsilon \, 
   \left\{ (\varepsilon - \mu) f(\varepsilon ) - \Omega (\varepsilon ) \right\} D(\varepsilon)  , 
\label{eq:chi_th_2}
\end{eqnarray}
where
\begin{eqnarray}
 \Omega (\varepsilon ) = - \int_\varepsilon^\infty d\varepsilon f(\varepsilon ) 
= - T \ln (1 + {\rm e}^{-\beta (\varepsilon - \mu)})  ,
\end{eqnarray}
assuming that $\varepsilon \, C(\varepsilon) \to 0$ as $\varepsilon \to -\infty$.
 We note that $\Omega (\varepsilon )$ is nothing but the grand-canonical free energy for fermions at energy $\varepsilon$.\cite{com10}  
 Since the first term in the brackets of Eq.~(\ref{eq:chi_th_2}) represents the (average) energy, 
$E(\varepsilon ) = \varepsilon f(\varepsilon )$, the terms in the brackets can be regarded as 
$E(\varepsilon ) - \mu f(\varepsilon) - \Omega (\varepsilon ) 
 = E(\varepsilon )- F (\varepsilon )  = TS(\varepsilon )$, 
where $F (\varepsilon ) \equiv \Omega (\varepsilon ) + \mu f(\varepsilon)$ is the corresponding Helmholtz free energy, and 
\begin{eqnarray}
 S (\varepsilon ) 
&=& \frac{\varepsilon - \mu}{T} f(\varepsilon ) + \ln (1 + {\rm e}^{-\beta (\varepsilon - \mu)}) 
\end{eqnarray}
is the entropy. 
 Thus we obtain the suggestive expression, 
\begin{eqnarray}
 \chi_{\rm th}^{\rm sea} 
&=& T \int_{-\infty}^\infty d\varepsilon \, S(\varepsilon ) D(\varepsilon)  . 
\label{eq:chi_th_3}
\end{eqnarray}
 Since the entropy behaves regularly and vanishes in the limit $T \to 0$, so does $\chi_{\rm th}^{\rm sea} /T$.\cite{com11}  
 The unphysical divergence has thus been removed. 

 If we define 
\begin{eqnarray}
 \Phi(T,\mu)
&=& \int_{-\infty}^\infty d\varepsilon \, \Omega (\varepsilon ) D(\varepsilon)  ,
\label{eq:tilde_M}
\end{eqnarray}
and note the relations, $f (\varepsilon ) = - \partial \Omega (\varepsilon ) /\partial \mu$  
and $S (\varepsilon ) = - \partial \Omega (\varepsilon ) /\partial T$, 
\begin{eqnarray}
 \chi_{\rm el}^{\rm sea} 
&=& - \frac{\partial}{\partial \mu} \Phi(T,\mu) , 
\label{eq:dP/dmu}
\\
 \chi_{\rm th}^{\rm sea} 
&=&  - T \frac{\partial}{\partial T} \Phi(T,\mu) ,  
\label{eq:dP/dT}
\end{eqnarray}
which looks very much like thermodynamic formulae. 
 Similar expressions are possible for the Fermi-surface terms as well.\cite{com_Psi} 
 A formula similar to Eq.~(\ref{eq:dP/dmu}) has been derived by St\v reda for the Fermi-sea term of the Hall conductivity. \cite{Streda82}

 The above considerations suggest the following prescription for the calculation of thermal response functions. 
 Given the electrical response functions, Eqs.~(\ref{eq:chi_el_surf_1}) and (\ref{eq:chi_el_general}), 
the thermal response functions, Eqs.~(\ref{eq:chi_th_surf_1}) and (\ref{eq:chi_th_3}), are obtained by the replacement, 
\begin{eqnarray}
 f (\varepsilon ) \to T S (\varepsilon ) . 
\label{eq:prescription}
\end{eqnarray}
 This prescription works for the Fermi-surface term as well, since 
$( - \partial f/\partial \varepsilon )$ is replaced by 
\begin{eqnarray}
 T \left( - \frac{\partial S}{\partial \varepsilon} \right) 
&=& (\varepsilon - \mu) \left( - \frac{\partial f}{\partial \varepsilon} \right) , 
\end{eqnarray}
which is identical with the $(\varepsilon - \mu)$-factor prescription for the Fermi-surface term, 
leading to Eq.~(\ref{eq:chi_th_surf_1}). 
 Although we did not derive this procedure from first principles, 
it suggests that a (fictitious) field that couples to the {\it entropy density} (times temperature), 
rather than to the energy (or heat) density, has more direct relevance for the problem.

\section{Summary}

 We presented a microscopic model calculation of spin torques 
induced by a temperature gradient in a conducting ferromagnet.
 Based on the observation that the Luttinger's prescription leads to an unphysical result, 
we recognized that the Einstein relation should be applied only to the non-equilibrium components. 
 We thus subtracted the equilibrium component from the Kubo formula before applying the Einstein relation.

 In the subtraction procedure, we noted 
(i) the modification of the torque formula by $\psi$ [Eq.~(\ref{eq:t_sd_1})], 
but (ii) the absence of a linear response to $\psi$ (not $\nabla \psi$); 
the latter reflects the adiabatic nature of the Kubo formula and the conservation of energy current (to which the field $\psi$ couples). 
 We suggest that a field which couples to the entropy density 
would directly lead to the desired results, although a formal proof is still necessary.

A general thermoelectric relation between thermal and electrical torques Eq.~(\ref{eq:t_th_result}) 
leads to a generalized Mott formula Eq.~(\ref{eq:t_T_2}) for sufficiently low temperatures. 
 When the dissipative correction ($\beta$-term) depends on energy, a new \lq\lq $\beta$-term'' beyond the simple thermoelectric effect (due to spin currents induced by temperature gradient) should be taken into account.

\acknowledgements

 HK would like to thank Erik van der Bijl for valuable discussion in the final stage of the present work. 
 This work was supported by Grants-in-Aid for Scientific Research (No.~21540336, 25400339 and 25247056) 
from the Japan Society for the Promotion of Science (JSPS). 

\appendix

\section{Linear Response to Gravitational Field}
\label{Linear Response}

 Here we summarize some formulae of the linear response to a \lq\lq gravitational'' field, $\psi$, 
which couples to the energy density of the system, as considered by Luttinger.\cite{Luttinger64} 

 To be specific, let us take 
$\psi ({\bm r},t) 
 = \psi_{\bm q} \, {\rm e}^{i({\bm q} \cdot {\bm r} -\omega t)}$. 
 Then the perturbation is described by 
\begin{eqnarray}
  H' = \psi_{\bm q} \, h (-{\bm q}) \, {\rm e}^{-i\omega t}
\end{eqnarray}
where $h({\bm q})$ is the Fourier component of the energy density $h(x)$. 
(In this paper, $h$ actually means $h - \mu n$, as stated just above Eq.~(\ref{eq:phenomenology}).) 
 To first order in $\psi$, the response of a physical quantity 
$\hat A$ is expressed as 
\begin{eqnarray}
 \langle \hat A \, \rangle_{\rm ne} 
= - K_0 ({\bm q}, \omega + i0) \, \psi_{\bm q} \, {\rm e}^{-i\omega t} .
\label{eq:A1}
\end{eqnarray}
 The response function $K_0 ({\bm q}, \omega + i0)$ is obtained from 
\begin{eqnarray}
 K_0 ({\bm q}, i\omega_\lambda ) 
 = \int_0^\beta d\tau \, {\rm e}^{i\omega_\lambda \tau} \, 
    \langle \, {\rm T}_\tau \, \hat A (\tau) 
            \, h(-{\bm q}) \, \rangle   
\label{eq:K_0}
\end{eqnarray}
by analytic continuation, $i\omega_\lambda \to \hbar\omega + i0$. 
 Let us introduce the heat-current operator ${\bm j}_Q$ 
by the continuity equation for the energy (measured from the chemical potential), 
\begin{eqnarray}
  \frac{\partial}{\partial t} \, h(x) + \nabla \!\cdot\! {\bm j}_Q = 0 . 
\label{eq:continuity}
\end{eqnarray}
 In the Fourier (${\bm q}$) and imaginary-time ($\tau$) representation, 
$\partial_\tau h(-{\bm q})= \hbar {\bm q} \!\cdot\! {\bm j}_Q (-{\bm q})$. 
 Using this in Eq.~(\ref{eq:K_0}) and doing a partial integration, we obtain 
\begin{eqnarray}
  K_0 ({\bm q}, i\omega_\lambda ) 
= \frac{\hbar q_i}{i\omega_\lambda} 
  [ K_i ({\bm q}, i\omega_\lambda ) - K_i ({\bm q},0)], 
\label{eq:K_0_1}
\end{eqnarray}
where 
\begin{eqnarray}
  K_i ({\bm q}, i\omega_\lambda ) 
= \int_0^\beta d\tau \, {\rm e}^{i\omega_\lambda \tau} \, 
    \langle \, {\rm T}_\tau \, \hat A (\tau) 
            \, j_{Q,i} (-{\bm q}) \, \rangle . 
\label{eq:K_i_0}
\end{eqnarray}
 The factor $iq_i$ in Eq.~(\ref{eq:K_0_1}) is combined with $\psi_{\bm q}$ 
in Eq.~(\ref{eq:A1}) to yield $\nabla \psi$. 
 When $\nabla \psi$ is uniform and static, we can take the limit 
${\bm q} \to {\bm 0}$ and $\omega \to 0$ in the coefficient 
(Eq.~(\ref{eq:K_i_0})) and obtain 
\begin{eqnarray}
&& \hskip -7mm 
 \langle \hat A \, \rangle_{\rm ne} 
 = \lim_{\omega \to 0}
    \frac{ K_i(\omega +i0) - K_i(0)}{i\omega} \, 
    \left( - \nabla_i \psi \right) , 
\label{eq:A2}
\\
&& \hskip -7mm 
 K_i (i\omega_\lambda ) 
 = \int_0^\beta d\tau \, {\rm e}^{i\omega_\lambda \tau} \, 
    \langle \, {\rm T}_\tau \, \hat A (\tau) 
            \, J_{Q,i} \, \rangle , 
\label{eq:K_i}
\end{eqnarray}
where ${\bm J}_Q \equiv {\bm j}_Q ({\bm q}={\bm 0})$ 
is the total heat current. 
 An explicit form of ${\bm j}_Q$ is given in Eq.~(\ref{eq:jQ}).

\section{Cancellation in the Interaction Picture}
\label{Cancellation}

 At the end of Sec.~IV, we showed that in the `Heisenberg' picture (defined for the full Hamiltonian, $H$) 
the heat-current vertex dffers from the charge current vertex only by the factor $i(\varepsilon_n + \omega_\lambda/2)$. 
. 
 Here we confirm this statement by a calculation 
based on the following explicit formula [Eq.~(\ref{eq:JQ})] for the heat current. 
 As seen below, due to many cancellations we are indeed left 
only with the first term of Eq.~(\ref{eq:KQ_2}).

 The explicit form of the total heat current operator 
(without invoking a time derivative as in Eq.~(\ref{eq:jQ})) is given by 
\begin{eqnarray}
&& \hskip -7mm 
 {\bm J}_Q
= \sum_{\bm k} {\bm v}_{\bm k} \, 
   c^\dagger_{\bm k} \, \hat \xi_{\bm k} \, c_{\bm k}^{\phantom{\dagger}} 
 - M \sum_{\bm k} {\bm v}_{\bm k} \, 
   c^\dagger_{{\bm k}+} \, ({\bm u}_{\bm q} \!\cdot {\bm \sigma}) \, 
   c_{{\bm k}-}^{\phantom{\dagger}} 
\nonumber \\
&& \hskip 0mm 
 + \frac{1}{2} \sum_{{\bm k},{\bm k}'} 
   ({\bm v}_{\bm k}+{\bm v}_{{\bm k}'}) \, 
   c^\dagger_{\bm k} \, V_{\rm imp} ({\bm k}-{\bm k}') \, 
   c_{{\bm k}'}^{\phantom{\dagger}} , 
\label{eq:JQ}
\end{eqnarray}
where ${\bm v}_{\bm k} = \hbar {\bm k}/m$, ${\bm k}\pm = {\bm k}\pm {\bm q}/2$,  
\begin{eqnarray}
&& \hskip -7mm 
 \hat \xi_{\bm k}
= \frac{\hbar^2{\bm k}^2}{2m} - M \sigma^z - \mu , 
\end{eqnarray}
and $V_{\rm imp} ({\bm k}-{\bm k}')$ is the Fourier transform of 
Eq.~(\ref{eq:V_imp}).

 Let us examine each contribution to $K_{ij}^{\alpha\beta}$; 
\begin{eqnarray}
&& \hskip -7mm 
 K_{ij}^{\alpha\beta} 
 = T \sum_n \sum_{{\bm k}} v_i v_j \, 
  (\varphi_1 + \varphi_2 + \varphi_3 + \varphi_2' + \varphi_3') .
\end{eqnarray}
 The contribution from the first term of Eq.~(\ref{eq:JQ}) 
is obtained by replacing the $v_i$-vertex 
in $\varphi_{ij}^{\alpha\beta}$ [Eq.~(\ref{eq:varphi_0})] as 
$G^+ v_i \, G \to G^+ v_i \, \hat \xi_{\bm k} G$. 
 Using the identity, $G^{-1} = i \varepsilon_n - \hat \xi_{\bm k} - \Sigma$, or 
\begin{eqnarray}
&& \hskip -7mm 
 \hat \xi_{\bm k} 
= 
i \left(\varepsilon_n + \frac{\omega_\lambda}{2} \right) 
 - \frac{1}{2} [(G^+)^{-1}+G^{-1}]
 - \frac{1}{2} (\Sigma^+ + \Sigma \, ) ,  
\nonumber \\
\end{eqnarray}
where $\Sigma$ is the self-energy, we have 
\begin{widetext}
\begin{eqnarray}
&& \hskip -12mm  
 \varphi_1 
= i (\varepsilon_n + \omega_\lambda /2 ) 
  \biggl\{ 
  {\rm tr} \left[ \sigma^\alpha G^+ G^+ \sigma^\beta G^+ G \right] 
 -{\rm tr} \left[ \sigma^\alpha G^+ G \sigma^\beta G G \right] \biggr\} ,
\nonumber \\ 
&& \hskip -12mm  
 \varphi_2 
= - \frac{1}{2} 
  \biggl\{ 
  {\rm tr} \left[ \sigma^\alpha G^+ G^+ \sigma^\beta (G^+ + G) \right] 
 -{\rm tr} \left[ \sigma^\alpha (G^+ + G) \, \sigma^\beta G G \right] 
  \biggr\} ,
\nonumber \\ 
&& \hskip -12mm  
 \varphi_3 
= - \frac{1}{2} 
  \biggl\{ 
  {\rm tr} \left[ \sigma^\alpha G^+ G^+ \sigma^\beta 
                  G^+ (\Sigma^+ + \Sigma \, ) \, G \right] 
- {\rm tr} \left[ \sigma^\alpha G^+ (\Sigma^+ + \Sigma \, ) \, 
                  G \, \sigma^\beta G G \right] \biggr\} . 
\end{eqnarray}
\end{widetext}

 In order to evaluate the contribution by the second term in Eq.~(\ref{eq:JQ}), 
we start from 
\begin{eqnarray}
&& \hskip -10mm  
 K_i^\alpha (i\omega_\lambda) 
= M u_{\bm q}^\beta \, T \sum_n \sum_{{\bm k}} v_i \, 
  {\rm tr} \left[ \sigma^\alpha G_{{\bm k}+{\bm q}/2}^+ 
                  \sigma^\beta G_{{\bm k}-{\bm q}/2} \right] 
\end{eqnarray}
and expand it with respect to $q_j$. 
 We obtain (Fig.~2 (a)) 
\begin{eqnarray}
&& \hskip -10mm  
 \varphi_2'
= \frac{1}{2} 
  \biggl\{ 
  {\rm tr} \left[ \sigma^\alpha G^+ G^+ \sigma^\beta G \right] 
 -{\rm tr} \left[ \sigma^\alpha G^+ \, \sigma^\beta G G \right] \biggr\} ,
\end{eqnarray}
which partly cancels $\varphi_2$; 
the remaining terms 
\begin{eqnarray}
&& \hskip -10mm  
 \varphi_2 + \varphi_2'
= \frac{1}{2} 
  \biggl\{ 
  {\rm tr} \left[ \sigma^\alpha G^+ G^+ \sigma^\beta G^+ \right] 
 -{\rm tr} \left[ \sigma^\alpha G \, \sigma^\beta G G \right] \biggr\} 
\end{eqnarray}
do not depend on $\omega_\lambda$ after summing over $\varepsilon_n$ 
and can be dropped. 
 This corresponds to the second term of Eq.~(\ref{eq:KQ_2}).

The contribution from the third term of Eq.~(\ref{eq:JQ}) is shown 
diagrammatically in Fig.~2 (b) and (c). 
 The diagrams of Fig.~2 (b) gives 
\begin{eqnarray}
&& \hskip -10mm  
 \varphi_3'
= \frac{1}{2} 
  \biggl\{ 
  {\rm tr} \left[ \sigma^\alpha G^+ G^+ 
                  \sigma^\beta G^+ (\Sigma^+ + \Sigma) G \right] 
\nonumber \\ 
&& \hskip 4mm 
-{\rm tr} \left[ \sigma^\alpha G^+ (\Sigma^+ + \Sigma) G \, 
                  \sigma^\beta G G \right] \biggr\} .
\end{eqnarray}
which cancels with $\varphi_3$. 
 The contribution of Fig.~2 (c) is $\sim {\cal O}(\gamma)$ and is disregarded. 

 For the diagrams including vertex corrections in Fig.~1, similar arguments hold. 

 As a result, we need to take into account only $\varphi_1$ 
(including vertex corrections), 
in accordance with the observation made around Eq.~(\ref{eq:KQ_2}).

\begin{figure}[t]
  \begin{center}
  \includegraphics[scale=0.32]{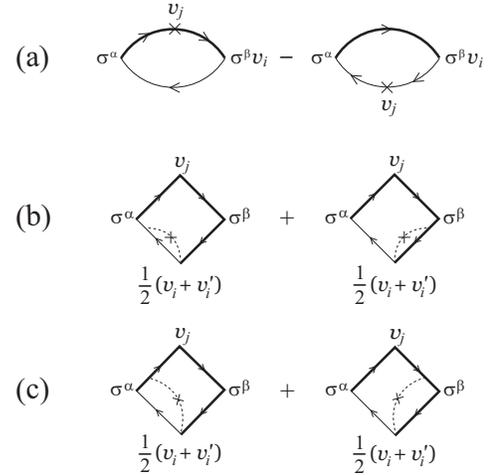}
  \vskip 0mm
  \end{center}
\caption{ 
 Diagrammatic expressions for $K_{Q,ij}^{\alpha\beta}$ 
calculated with the second term (a) and the third term (b,c) 
of the heat-current operator, Eq.~(\ref{eq:JQ}). 
}
\label{fig:K_correction}
\end{figure}

\section{Equilibrium Exchange Torque}
\label{Exchange}

 Here we calculate the equilibrium exchange torque, Eq.~(\ref{eq:t_eq}), to show that it indeed has the same coefficient $c$ [Eq.~(\ref{eq:c})] as the problematic term, Eq.~(\ref{eq:unwanted}). 
 In the presence of a static magnetization texture, Eq.~(\ref{eq:u}), the equilibrium spin density to the first order in ${\bm u}_{\bm q}$ reads   
\begin{eqnarray}
 \langle \hat \sigma_\perp^\alpha ({\bm q}) \rangle_{\rm eq} 
= M K^{\alpha\beta} ({\bm q}) u_{\bm q}^\beta  , 
\end{eqnarray}
where 
\begin{eqnarray}
 K^{\alpha\beta} ({\bm q})
&=& - T\sum_n \sum_{\bm k} {\rm tr} [ \sigma^\alpha G_{{\bm k} + {\bm q}} (i\varepsilon_n) \sigma^\beta G_{\bm k} (i\varepsilon_n)]  
\nonumber \\
&=&  K^{\alpha\beta} ({\bm 0}) + K_{ij}^{\alpha\beta} q_i q_j  + {\cal O}(q^4)  . 
\end{eqnarray}
 In the second line, we expanded $K^{\alpha\beta} ({\bm q})$ with respect to ${\bm q}$ 
with coefficients $K^{\alpha\beta} ({\bm 0}) = (\rho_{\rm s}/M) \delta^{\alpha\beta}$, 
where $\rho_{\rm s}=n_\uparrow - n_\downarrow$ is the conduction electron spin polarization for uniform ${\bm n}$, and 
\begin{eqnarray}
 K_{ij}^{\alpha\beta} 
&=& \frac{1}{2} T\sum_n \sum_{\bm k} v_i v_j {\rm tr} [ \sigma^\alpha GG \sigma^\beta GG]  
\nonumber \\
&=&  \delta^{\alpha\beta} \, T\sum_n \sum_{\bm k} v_i v_j  (G_\uparrow G_\downarrow )^2   
\end{eqnarray}
with $G \equiv G_{\bm k}(i\varepsilon_n)$ and  $G_\sigma \equiv G_{{\bm k}\sigma}(i\varepsilon_n)$. 
 Analytic continuation leads to 
\begin{eqnarray}
 K_{ij}^{\alpha\beta} 
&=&  - \frac{1}{\pi} \delta^{\alpha\beta} \int_{-\infty}^\infty d\varepsilon f(\varepsilon) 
   \sum_{\bm k} v_i v_j  
   {\rm Im} [G_\uparrow^{\rm R} (\varepsilon) G_\downarrow^{\rm R} (\varepsilon)]^2 
\nonumber \\
&=&  - \delta^{\alpha\beta} \frac{1}{M^2} \int_{-\infty}^\infty d\varepsilon f(\varepsilon) C(\varepsilon) 
\nonumber \\
&=&  - \frac{c}{M^2} \delta^{\alpha\beta} , 
\end{eqnarray}
where we used Eqs.~(\ref{eq:C}) and (\ref{eq:c}). 
 This gives the spin density, 
$\langle \hat {\bm \sigma} \rangle_{\rm eq} 
= \rho_{\rm s} \hat z + \langle \hat {\bm \sigma}_\perp ({\bm q}) \rangle_{\rm eq}  
= \rho_{\rm s} (\hat z + {\bm u}) +  (c/M) \, \nabla^2 {\bm u} 
= \rho_{\rm s} {\bm n} +  (c/M) \, \nabla^2 {\bm n} $, 
and the torque, Eq.~(\ref{eq:t_eq}).

\section{Response to Scalar Potentials}
\label{Scalar}

Here we directly calculate the linear response to the scalar potentials of electric ($\phi$) and gravitational 
($\psi$) fields. 
 This confirms our assertion that no terms proportional to $\psi$ arise (next to those with $\partial_i \psi$), 
which is crucial for the procedure proposed in Sec.~VI. 
 It also serves as a check of Eqs.~(\ref{eq:A2}) and (\ref{eq:K_i}).

 The linear response of the $s$-electron spin density to $\phi$ or $\psi$ may be expressed as 
\begin{eqnarray}
&& \hskip -6mm 
 \langle \hat {\bm \sigma}_\perp \rangle_\phi  
= -e \, (A_\phi -i\omega \, B_\phi )/M , 
\label{eq_app:s_phi}
\\
&& \hskip -6mm 
 \langle \hat {\bm \sigma}_\perp \rangle_\psi  
= (A_\psi -i\omega \, B_\psi )/M , 
\label{eq_app:s_psi}
\end{eqnarray} 
respectively, retaining the terms up to first order in $\omega$, {\it i.e.}, 
the frequency of $\phi$ or $\psi$.   
 The coefficients are   
\begin{eqnarray}
&& \hskip -5mm 
 A_\phi = - C_0 \, \partial_i [(\partial_i {\bm n}) \, \phi ] , 
\label{eq_app:A_phi} 
\\
&& \hskip -5mm 
 B_\phi 
= \frac{ C_0 \, (\nabla^2 {\bm n}) \, \phi 
       + [B_0 \, \partial_i {\bm n} - A_0 ({\bm n} \times \partial_i {\bm n}) ] \, \partial_i \phi}{-i\omega} ,
\label{eq_app:B_phi} 
\\
&& \hskip -5mm 
 A_\psi = (c - C_1) \, \partial_i [(\partial_i {\bm n}) \, \psi ] 
         - c \, (\nabla^2 {\bm n}) \, \psi , 
\label{eq_app:A_psi} 
\\
&& \hskip -5mm 
 B_\psi 
= \frac{ C_1 \, (\nabla^2 {\bm n}) \, \psi 
        + [ B_1 \, \partial_i {\bm n} - A_1 ({\bm n} \times \partial_i {\bm n}) ] \, \partial_i \psi}{-i\omega} ,
\label{eq_app:B_psi}  
\end{eqnarray}
where $A_n, B_n, C_n$ and $c$ are given by Eqs.~(\ref{eq:An}) and (\ref{eq:c}). 
 The second term in Eq.~(\ref{eq_app:A_psi}) 
is a correction similar to the second term in Eq.~(\ref{eq:KQ_2}) treating the heat vertex by the factor $i(\varepsilon_n + \omega_\lambda /2)$. 
 Each factor $(-i\omega)^{-1}$ in Eqs.~(\ref{eq_app:B_phi}) and (\ref{eq_app:B_psi}) reflects conservation 
of electron number and energy, respectively, and comes from ladder-type vertex correction.\cite{SK11,Hosono12} 
 Therefore, even in the static limit, $\omega \to 0$, 
the $B_\phi$- and $B_\psi$-terms survive in Eqs.~(\ref{eq_app:s_phi}) and (\ref{eq_app:s_psi}) and lead to
\begin{eqnarray}
&& \hskip -10mm 
 M \langle \hat {\bm \sigma}_\perp \rangle_\phi 
= -e [ (B_0-C_0) \, \partial_i {\bm n} -A_0 ({\bm n} \times \partial_i {\bm n}) ] \, \partial_i \phi ,
\label{eq_app:s_phi_1}
\\
&& \hskip -10mm 
 M \langle \hat {\bm \sigma}_\perp \rangle_\psi 
= [ (B_1-C_1+c) \, \partial_i {\bm n} -A_1 ({\bm n} \times \partial_i {\bm n}) ] \, \partial_i \psi .
\label{eq_app:s_psi_1}
\end{eqnarray}
 Note that the terms proportional to $\phi$ or $\psi$ (but not $\partial_i \phi$ or $\partial_i \psi$) 
cancel exactly, which reflects the adiabatic nature of the Kubo formula (see Appendix \ref{Thermodynamics}),  
and is crucial for the procedure described in Sec.~VI.  
 Torques obtained from Eqs.~(\ref{eq_app:s_phi_1}) and (\ref{eq_app:s_psi_1}) agree  
with Eqs.~(\ref{eq:t_phi}) and (\ref{eq:t_psi}), 
confirming the validity of Eqs.~(\ref{eq:A2}) and (\ref{eq:K_i}).

\section{Response to Static and Uniform Scalar Potentials}
\label{Thermodynamics}

 In this Appendix, we consider $\phi$ and $\psi$ that are static and uniform. 
 The response to such potentials can be compared with equilibrium theory.

 The perturbation is described by the Hamiltonian 
\begin{eqnarray}
  H' =  - e N \phi + K \, \psi  , 
\label{eq_th:h'}
\end{eqnarray}
where $K=H - \mu N$, $N$ is the total number of electrons, and $H$ is the Hamiltonian of the (unperturbed) system. 
(We neglect the nonlinear perturbation proportional to $\phi \psi$.)
 Let us consider the adiabatic and isothermal response of a physical quantity $\hat A$, 
\begin{eqnarray}
 \delta \langle \hat A \, \rangle^{\rm ad} = e \chi_N^{\rm R} (0) \, \phi - \chi_K^{\rm R} (0) \, \psi  ,
\label{eq_th:A1}
\\
 \delta \langle \hat A \, \rangle^T = e \chi_N^T (0) \, \phi - \chi_K^T (0) \, \psi  ,
\label{eq_th:AT1}
\end{eqnarray}
respectively. 
 The response functions are given by the static limit of 
\begin{eqnarray}
&& \hskip -7mm 
 \chi_B^{\rm R} (\omega) 
= \frac{i}{\hbar} \int_0^\infty dt \, {\rm e}^{i(\omega + i\eta) t} \, 
                  \langle \, [\, \hat A (t) , \, \hat B \, ] \, \rangle  ,
\label{eq_th:KR_0}
\\
&& \hskip -7mm 
 \chi_B^T (i\omega_\lambda ) 
= \int_0^\beta d\tau \, {\rm e}^{i\omega_\lambda \tau} \, 
    \langle \, {\rm T}_\tau \, \hat A (\tau) \, \Delta \hat B \, \rangle ,
\label{eq_th:K_0}
\end{eqnarray}
where $\Delta \hat B = \hat B - \langle \, \hat B \, \rangle$, 
with $\hat B=N$ or $K$.\cite{Kubo83} 
 Since $\hat B$ commutes with $K$, we have 
\begin{eqnarray}
&& \hskip -7mm 
 \chi_B^{\rm R} (\omega) = 0  , 
\\
&& \hskip -7mm 
 \chi_B^T (i\omega_\lambda) 
= \beta \, \langle \, \Delta \hat A \, \Delta \hat B \, \rangle \, \delta_{\lambda ,0} .  
\label{eq_th:<AdB>}
\end{eqnarray}
 Therefore, the adiabatic response vanishes,  
\begin{eqnarray}
 \delta \langle \hat A \, \rangle^{\rm ad} = 0 .
\label{eq_th:dA_ad}
\end{eqnarray} 
 The Kubo formula corresponds to this case.\cite{Kubo83} 
 The isothermal response (\ref{eq_th:<AdB>}) can be expressed by the thermodynamic formula,  
\begin{eqnarray}
&& \hskip -7mm 
 \chi_N^T (0) 
= \frac{\partial}{\partial \mu} \langle \, \hat A \, \rangle ,
\label{eq_th:<AdN>}
\\
&& \hskip -7mm 
 \chi_K^T (0) 
= - \beta \frac{\partial}{\partial \beta} \langle \, \hat A \, \rangle 
= T \frac{\partial}{\partial T} \langle \, \hat A \, \rangle ,
\label{eq_th:<AdH>}
\end{eqnarray}
for $\hat B=N$ and $K$, respectively, leading to 
\begin{eqnarray}
 \delta \langle \hat A \, \rangle^T  
= e\phi \,  \frac{\partial}{\partial \mu} \langle \, \hat A \, \rangle 
 - \psi \, T \frac{\partial}{\partial T} \langle \, \hat A \, \rangle .
\label{eq_th:dA_M}
\end{eqnarray}
 This is natural since 
${\rm e}^{-\beta (K + H')} = {\rm e}^{-\beta [(1+\psi )K - e\phi N]}$  
is nothing but ${\rm e}^{-\beta K} = {\rm e}^{-\beta (H-\mu N)}$ with $\beta$ and $\mu$ 
modified by $\delta \beta = \beta \, \psi$ and $\delta \mu = e \phi$, respectively.

 Here we are interested in $\hat A = \hat \sigma^\alpha_\perp$ with equilibrium value (see Appendix C) 
\begin{eqnarray}
 \langle \hat {\bm \sigma}_\perp \rangle 
= c \, \nabla^2 {\bm n} \, /M , 
\label{eq_th:s_eq}
\end{eqnarray}
where $c$ is given by Eq.~(\ref{eq:c}). 
 Since $\partial c / \partial \mu = C_0$ and $T (\partial c / \partial T) = C_1$, 
the isothermal response is given by 
\begin{eqnarray}
 \delta \langle \hat {\bm \sigma}_\perp \, \rangle^T
= ( e\phi \, C_0- \psi \, C_1 \, ) \, \nabla^2 {\bm n} \, /M .
\label{eq_th:ds_T}
\end{eqnarray}
 The susceptibilities read $\chi_N^T = C_0 \nabla^2 {\bm n} \, /M$ and $\chi_K^T = C_1 \nabla^2 {\bm n} \, /M$. 
 The adiabatic susceptibilities, $\chi_N^{\rm ad}$ and $\chi_K^{\rm ad}$, are obtained by subtracting the corrections 
due to changes in $T$ and $\mu$,\cite{Kubo83} giving $\chi_N^{\rm ad}=\chi_K^{\rm ad}=0$, consistent with Eq.~(\ref{eq_th:dA_ad}). 

 We recognize these isothermal components (\ref{eq_th:ds_T}) in Eqs.~(\ref{eq_app:A_phi}) and (\ref{eq_app:A_psi}), 
which are eventually canceled by the corresponding terms in Eqs.~(\ref{eq_app:B_phi}) and (\ref{eq_app:B_psi}), 
resulting in a vanishing adiabatic response to static and uniform $\phi$ and $\psi$.


\begin{thebibliography}{00}
\bibitem{com_sd} 
Equivalent relations can be written for the Stoner model.
\bibitem{com:beta} The same letter ($\beta$) is also used for the coefficient of the $\beta$-term 
as well as for the spin index (such as $\varepsilon^{\alpha\beta}$), but we hope no confusion
will arise.
\bibitem{com4} 
 In Eq.~(\ref{eq:varphi_0}), $\sigma^\beta$ can be moved to the position adjacent to $\sigma^\alpha$ 
by reversing the spin, $\sigma^z \to - \sigma^z$, in $G$ and make use of  
$\sigma^\alpha \sigma^\beta 
 = \delta^{\alpha \beta} + i\sigma^z \varepsilon^{\alpha \beta}$. 
 This dependence on $\alpha$ and $\beta$ was exploited 
in Eqs.~(\ref{eq:K_i_expansion}) and (\ref{eq:K_Qi_expansion}).
\bibitem{com_sea} 
 Since $f(\varepsilon)$ is multiplied by a total derivative $\partial_\varepsilon C(\varepsilon)$ 
in the Fermi-sea term in $\tilde b$ [second term in Eq.~(\ref{eq:b})],  
this term can be transformed into a Fermi-surface term by partial integration. 
 However, this is specific to the present case; in general such a total-derivative form does not 
appear at the formal Green function level, see Ref.~\onlinecite{vdBijl}.
\bibitem{com:sigma_s} 
 If we neglect magnetic impurities, $\tau_\sigma^{-1} \propto \nu_\sigma$ [Eq.~(\ref{eq:gamma})] and thus 
$\sigma_{\rm s} \propto \sum_\sigma \sigma n_\sigma \tau_\sigma 
\propto \sum_\sigma \sigma \varepsilon_{{\rm F} \sigma} = 2M$ does not depend on $\varepsilon_{\rm F}$. 
 Magnetic impurities give rise to $\varepsilon_{\rm F}$-dependent terms in $\sigma_{\rm s}$ 
with a relative magnitude $\sim (n_{\rm s} u_{\rm s}^2 S_{\rm imp}^2 / n_{\rm i} u_{\rm i}^2)^3$.
\bibitem{com10} 
Note that $\varepsilon$ is the off-shell energy and does not need to correspond to real energy levels of the system.
\bibitem{com11} 
 If $D(\varepsilon )$ were a (single-particle) density of states, 
$\chi_{\rm el}^{\rm sea}$, $\chi_{\rm th}^{\rm sea}/T$ and $\Phi$ [defined in Eq.~(\ref{eq:tilde_M})] 
would represent the number, entropy and the grand-canonical free energy, respectively. 
 The \lq unsubtracted' $\chi_{\rm th}^{\rm sea}$ [first term of $\chi_{\rm th}^{\rm sea}$ in Eq.~(\ref{eq:chi_th_general})] then corresponds to an energy, 
which does not vanish at zero temperature because of Fermi degeneracy.
\bibitem{com_Psi} 
 With  
$ \Psi \equiv \int_{-\infty}^\infty d\varepsilon \, f (\varepsilon ) B(\varepsilon) $, 
we have $\chi_{\rm el}^{\rm surface} =  \partial \Psi / \partial \mu$ and 
$\chi_{\rm th}^{\rm surface} = T (\partial \Psi / \partial T)$.
\end{thebibliography}
\end{document}